\newcommand{\CLASSINPUTtoptextmargin}{0.77in}
\newcommand{\CLASSINPUTbottomtextmargin}{1.02in}

\documentclass[10pt,conference]{IEEEtran}
\usepackage{cite}
\usepackage{amsmath,amssymb,amsfonts}
\usepackage{algorithmic}
\usepackage{graphicx}
\usepackage{textcomp}
\usepackage{xcolor}
\usepackage{glossaries}
\usepackage{siunitx}
\usepackage{multirow}
\usepackage{subfig}
\usepackage[hidelinks]{hyperref}
\usepackage{textpos}

\newacronym{XR}{XR}{Extended Reality}
\newacronym{6DoF}{6DoF}{6 Degrees of Freedom}
\newacronym{AP}{AP}{Access Point}
\newacronym{SLS}{SLS}{Sector-Level Sweep}
\newacronym{DTI}{DTI}{Data Transmission Interval}
\newacronym{AWV}{AWV}{Antenna Weight Vector}
\newacronym{TXSS}{TXSS}{Transmit Sector Sweep}
\newacronym{A-BFT}{A-BFT}{Association - Beamforming Training}
\newacronym{HMD}{HMD}{Head-Mounted Device}
\newacronym{BHI}{BHI}{Beacon Header Interval}
\newacronym{BI}{BI}{Beacon Interval}
\newacronym{MCS}{MCS}{Modulation and Coding Scheme}
\newacronym{QoE}{QoE}{Quality of Experience}
\newacronym{CDF}{CDF}{Cumulative Distribution Function}
\newacronym{mmWave}{mmWave}{Millimeter-Wave}
\newacronym{BRP}{BRP}{Beam Refinement Phase}
\newacronym{BTI}{BTI}{Beacon Transmission Interval}
\newacronym{LoS}{LoS}{Line of Sight}
\newacronym{MU-MIMO}{MU-MIMO}{Multi-User Multiple-Input Multiple-Output}
\def\BibTeX{{\rm B\kern-.05em{\sc i\kern-.025em b}\kern-.08em
    T\kern-.1667em\lower.7ex\hbox{E}\kern-.125emX}}
\begin{document}

\title{Multi-Gigabit Interactive Extended Reality over Millimeter-Wave: An End-to-End System Approach}

\author{\IEEEauthorblockN{Jakob Struye}
\IEEEauthorblockA{\textit{IDLab - Department of Computer Science} \\
\textit{University of Antwerp - imec}\\
Antwerp, Belgium \\
jakob.struye@uantwerpen.be}
\and
\IEEEauthorblockN{Filip Lemic}
\IEEEauthorblockA{\textit{i2Cat Foundation} \\
Barcelona, Spain \\
filip.lemic@i2cat.net}
\and
\IEEEauthorblockN{Jeroen Famaey}
\IEEEauthorblockA{\textit{IDLab - Department of Computer Science} \\
\textit{University of Antwerp - imec}\\
Antwerp, Belgium \\
jeroen.famaey@uantwerpen.be}
}

\maketitle

\begin{abstract}
Achieving high-quality wireless interactive Extended Reality (XR) will require multi-gigabit throughput at extremely low latency. The Millimeter-Wave (mmWave) frequency bands, between 24 and \SI{300}{\giga\hertz}, can achieve such extreme performance. However, maintaining a consistently high Quality of Experience with highly mobile users is challenging, as mmWave communications are inherently directional. In this work, we present and evaluate an end-to-end approach to such a mmWave-based mobile XR system. We perform a highly realistic simulation of the system, incorporating accurate XR data traffic, detailed mmWave propagation models and actual user motion. We evaluate the impact of the beamforming strategy and frequency on the overall performance. In addition, we provide the first system-level evaluation of the CoVRage algorithm, a proactive and spatially aware user-side beamforming approach designed specifically for highly mobile XR environments.
\end{abstract}

\section{Introduction}\label{sec:intro}
\begin{textblock}{180}(0,6.45)
    \begin{tiny}
    \copyright\ 2024 IEEE. Personal use of this material is permitted. Permission from IEEE must be obtained for all other uses,\\
    \vspace{-6mm}\\
    in any current or future media, including reprinting/republishing this material for advertising or promotional purposes,\\
    \vspace{-6mm}\\
    creating new collective works, for resale or redistribution to servers or lists, or reuse of any copyrighted component\\
    \vspace{-6mm}\\
    of this work in other works.
    \end{tiny}
    \end{textblock}
\gls{XR} refers to technologies in which users are exposed to an immersive virtual world (Virtual Reality) or hybrid virtual/real world (Augmented and Mixed Reality) through a wearable device, such as a \gls{HMD}. From a networking perspective, \textit{interactive} \gls{XR} is by far the most challenging branch of \gls{XR} to achieve a consistently high \gls{QoE} in. As its content is generated in real-time, timely delivery requires extremely high throughput and low latency. The most well-known application is gaming~\cite{vrgaming1, vrgaming2}, with other applications, such as tele-operation and remote presence (e.g., meetings) also emerging~\cite{telepresence}. Interactive \gls{XR} is expected to benefit from wireless communications in the \gls{mmWave} range, comprising 24 to \SI{300}{\giga\hertz}, as lower frequencies cannot fulfill the multi-gigabit throughput requirements~\cite{mmwavevroverview}. However, utilizing these frequencies comes with a set of challenges. Most notably, the higher path and penetration losses make establishing high-quality links with \gls{mmWave} highly challenging. The main mitigation technique is \textit{beamforming}, meaning energy is directionally focused towards a peer, enabling high-gain links with a modest energy budget. This is especially challenging in \gls{XR} scenarios, as directions change rapidly with highly mobile users. 
Furthermore, overhead periods dedicated to control communications are in the order of milliseconds with \gls{mmWave}, making their impact on timely delivery of real-time traffic, such as \gls{XR} content, non-negligible~\cite{towards}.

While \gls{mmWave}-based \gls{XR} systems have frequently been proposed and evaluated over the past years, uncertainty about the upper performance limits in multi-gigabit, high-mobility deployments remains. In theory, \gls{mmWave} Wi-Fi can offer up to \SI{8}{Gbps} with a single channel and physical path, but the fluctuating link quality along with the bursty nature and extremely strict latency requirements of the traffic, make multi-gigabit systems challenging to achieve in practice~\cite{towards}. 

In this work, we therefore present and evaluate an end-to-end system approach for high-\gls{QoE} \gls{mmWave}-based interactive mobile \gls{XR}. Our system uses IEEE 802.11ay, the most recent \gls{mmWave} amendment to the Wi-Fi specification, and is evaluated through a simulation of the networking stack. The simulation includes realistic signal propagation through ray tracing, accurate \gls{XR} data streams and real-world user motion traces. Through an extensive experimentation, we show how a \SI{7}{Gbps} \gls{XR} stream can be delivered at high reliability to a highly mobile user. Achieving this requires rapid beamforming at the user side, as angles of arrival change rapidly during rapid rotational motion. To this end, we integrate the CoVRage algorithm, allowing for proactive beamforming based on predicted rotations, as presented in our previous work~\cite{covrage}. Based on these results, we provide several generic guidelines for achieving \gls{mmWave}-based interactive \gls{XR}.

Overall, this work presents the following contributions:
\begin{itemize}
    \item An end-to-end system approach to interactive \gls{XR} over \gls{mmWave}, targeting consistently high \gls{QoE} for highly mobile users
    \item A simulation-based analysis of the system's performance, incorporating realistic traffic patterns, real-world \gls{6DoF} user motion, multi-gigabit data rates, highly realistic channel and protocol behavior and metrics representative of actual \gls{XR} performance
    \item Open-source extensions to the IEEE 802.11ay module for the \texttt{ns-3} network simulator, including extensions to beamforming functionality and a codebook generator
    \item The first full-system evaluation of CoVRage, highlighting its advantages
    \item A set of generic guidelines and insights for \gls{mmWave}-based interactive \gls{XR}
\end{itemize}
The remainder of this paper is structured as follows. Sections \ref{sec:bg} and \ref{sec:rw} provide an overview of IEEE 802.11's \gls{mmWave} functionality and of related work on \gls{mmWave} \gls{XR} evaluations. Section \ref{sec:setup} contains an overview of the system. Then, Sections \ref{sec:exp} and \ref{sec:eval} present the simulation experiments and their results, with Section \ref{sec:conclusions} concluding this paper.
\section{Background} \label{sec:bg}
Given the immense requirements wireless interactive \gls{XR} places on the network, solutions based on the commonly used sub-\SI{6}{\giga\hertz} bands are not expected to provide sufficiently high performance. These bands are often highly congested, due to their ubiquitous usage along with limited bandwidth. As such, \gls{mmWave} communications, comprising 24 to \SI{300}{\giga\hertz}, are considered a necessity for achieving high-\gls{QoE} future interactive \gls{XR}. 
\subsection{Millimeter-Wave}
The two main current wireless communications standards, IEEE 802.11 and 5G, each provide dedicated \gls{mmWave} functionality. IEEE 802.11's two \gls{mmWave}-focused amendments, IEEE 802.11ad and the more recent upgrade IEEE 802.11ay, are especially focused on extremely-high-performance indoors communications, making this technology an ideal fit for indoors wireless interactive \gls{XR}~\cite{social}. A large part of the \gls{mmWave} functionality addresses challenges which are less prevalent in sub-\SI{6}{\giga\hertz} bands, stemming from \gls{mmWave}'s inherently high path loss and low penetration and reflection capabilities. To ensure sufficient signal strength over even modest distances, energy must be focused in some intended direction through a process called \textit{beamforming}~\cite{11ad}. To achieve this, \gls{mmWave} devices are equipped with \textit{antenna arrays}, in which each element transmits or receives a signal with a different phase shift applied to it. Given a carefully selected set of these phase shifts, called an \gls{AWV}, signals will phase-align and therefore interfere constructively in this intended direction, but destructively in others. To determine an appropriate \gls{AWV}, \textit{codebooks} are often used. These define a number of \textit{sectors}, each having an \gls{AWV} for communications in a different direction. By exhaustively testing every sector in a \textit{sector sweep}, the one providing the highest signal strength can be found. The \gls{mmWave} Wi-Fi specification defines two processes for \gls{AWV} selection: the \gls{SLS} and the \gls{BRP}. During the \gls{SLS}, which allows for initial association, one side will iterate through all its sectors while the other side deploys a \textit{quasi-omni} \gls{AWV}, which aims to have a similar signal strength in all directions. Once completed, the two sides switch roles. Then, the \gls{BRP} allows for further optimization of the \gls{AWV} through multiple mechanisms, some optional to implement. As beamforming can occur both when transmitting and when receiving, the nodes choose whether to train either their transmit or their receive \gls{AWV} during the \gls{SLS}. Commonly, both will train their transmit \gls{AWV} then, as receive \gls{AWV} training can be postponed until the \gls{BRP}. 

\begin{figure}[!t]
    \centering
    \includegraphics[width=\linewidth]{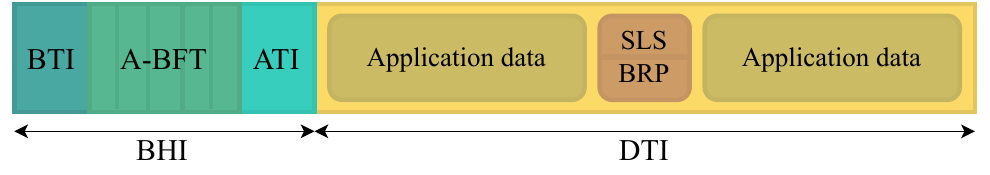}
    \caption{The Beacon Interval}
    \label{fig:biconfig}
  \end{figure}

\Gls{mmWave} Wi-Fi provides a dedicated period for \gls{AP} discovery and association in its schedule, shown in Fig.~\ref{fig:biconfig}. Once every \gls{BI} (\SI{102.4}{\milli\second} by default), a \gls{BHI} period is scheduled. The \gls{BHI} starts with a \gls{BTI}, during which the \gls{AP} transmits a \textit{beacon} on each of its sectors iteratively, informing nodes of its existence and capabilities. If nodes receive these beacons using their quasi-omni \gls{AWV}, this doubles as the first half of the \gls{SLS}. The second half can then be performed during the next sub-phase of the \gls{BHI}, the \gls{A-BFT}. This consists of several \textit{slots} during which nodes can choose to initiate the second half of the \gls{SLS}. Following a successful \gls{SLS}, a follow-up \gls{BRP} may be scheduled either within that \gls{A-BFT}, or as part of the \gls{DTI}, the phase following the \gls{BHI}. Repeat \glspl{SLS} with already associated nodes, for example when node mobility threatens to degrade a link, can also be scheduled in this \gls{DTI}. The \gls{DTI} is intended mainly for application-level traffic, but parts of it can be reserved for beamforming overhead when needed. In contrast, no application-level traffic may be scheduled during the \gls{BHI}, which occurs at the start of every \gls{BI} regardless of current demand for beamforming.
  
\subsection{CoVRage}
Beamforming as described above is an inherently reactive process. In high-mobility scenarios, such as wireless interactive \gls{XR}, where a user's instantaneous rotational speed may reach hundreds of degrees per second~\cite{followingTheShadow}, this does not suffice to maintain a stable wireless link. Rotational motion can cause rapid beam misalignment, with occasional beamforming not being responsive enough to adapt in time. To alleviate this, we introduced the \gls{HMD}-side \textit{CoVRage} beamforming algorithm in our previous work~\cite{covrage}. The algorithm receives an estimated current and an upcoming predicted pose as inputs. From this, it interpolates the rotational trajectory on which the \gls{AP} will travel, taking the \gls{HMD} as point of reference. The \gls{HMD} antenna array is then logically divided into sub-arrays, with each sub-array aiming a \textit{sub-beam} somewhere along the trajectory, while minimizing destructive interference between adjacent sub-beams. As such, CoVRage can synthesize irregularly shaped beams that will offer a consistently high gain towards the \gls{AP} during an expected upcoming rotation. This results in a very rapid, proactive beamforming approach, bypassing sector-based searches entirely, instead relying on modern \glspl{HMD}' accurate pose estimation and prediction functionality. The algorithm was thoroughly evaluated through PHY-level simulations, but was not yet integrated and evaluated as part of an end-to-end \gls{XR} system. 
\begin{figure*}[!t]
    \centering
    \includegraphics[width=\linewidth]{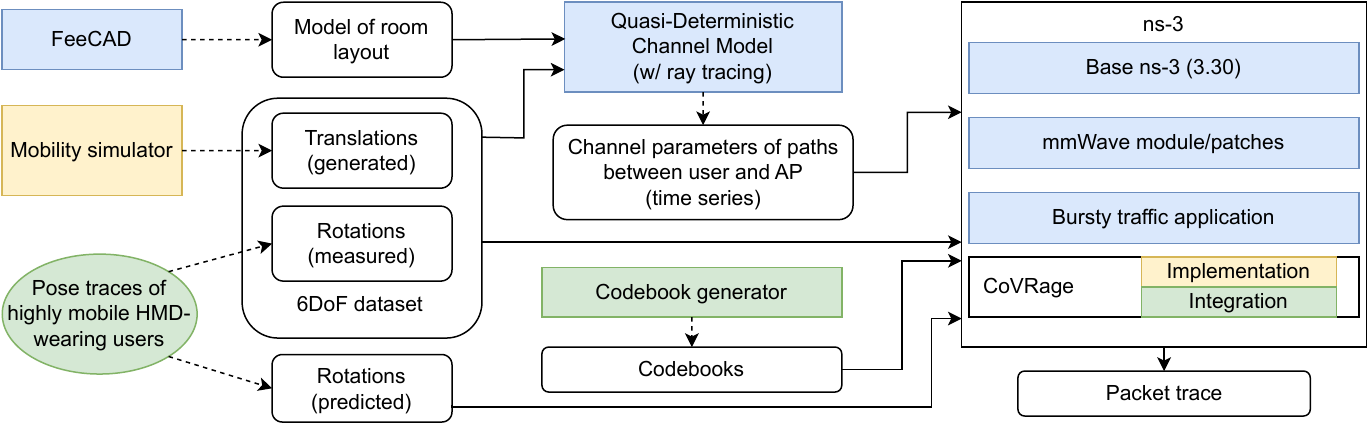}
    \caption{Overview of the simulation's components. Rectangles denote software, rounded rectangles denote datasets, and ovals denote user experiments. Blue indicates existing open-source software which remained largely unaltered, yellow indicates software by this paper's authors previously presented in other work, and green indicates new software (or experiments) implemented (or performed) especially for this work. All datasets were gathered for this work specifically.}
    \label{fig:overview}
    \end{figure*}
\section{Related Work}\label{sec:rw}
Some evaluations of \gls{XR} over \gls{mmWave} have been performed, with the performance of the networking aspect being simulated, modelled analytically, or measured empirically using experimental or off-the-shelf hardware. One work investigates the user capacity of such a system with users requiring up to \SI{45}{Mbps}, through a proprietary 5G simulator~\cite{5gsimNokia}. Another work transmits real 3D video and incorporates measured \gls{6DoF} user motion into an \texttt{ns-3} simulation using the \gls{mmWave} Wi-Fi module and quasi-deterministic channel model, reporting a high-level metric on user immersion fidelity~\cite{mmwavevsfso}. Using a similar framework, the authors in \cite{mumimoeval} investigate the capabilities of IEEE 802.11ay with \gls{MU-MIMO} at multi-gigabit data rates. They consider multiple user orientations, but do not consider performance \textit{during} motion. Next, in an evaluation on the value of mobile edge computing, the packet latency is reported using the 5G module for \texttt{ns-3}~\cite{mec}. While the appropriate video tiles to transmit were selected based on head rotations, it is unclear if these rotations were also considered within the network simulation. The next work compares IEEE 802.11ay link-level throughput with both \texttt{ns-3} and real hardware, finding the two to be similar~\cite{mitras}. Similarly, another work compares raw, UDP and TCP throughputs as a function of distance for both the \texttt{ns-3} 5G module and IEEE 802.11ad-enabled laptops, with the two again being similar~\cite{dellcomp}. Another work compares low-level performance of older IEEE802.11ad dongles to a simple, custom simulation model~\cite{dongle}. In terms of hardware-only evaluations, one work investigates throughput reduction caused by mobility-induced blockage when transmitting \gls{XR} content using off-the-shelf IEEE 802.11ad routers~\cite{talononbot}. One final work presents a detailed model encompassing user mobility, \gls{XR} traffic patterns, channel behavior, transmission schedules and signaling overheads~\cite{social}. They report per-packet losses for different maximal latencies and throughputs (max. \SI{200}{Mbps}).

Overall, none of the existing works cover all the functionality we incorporate into our system, outlined in our contributions in Section~\ref{sec:intro}.
\section{End-to-End Mobile XR System}\label{sec:setup}
In this work, we present and evaluate a full end-to-end wireless mobile \gls{XR} system. We present its design, alongside its implementation in simulation.
\subsection{Design}
The interactive \gls{XR} system is deployed in a large room, with a central ceiling-mounted IEEE 802.11ad \gls{AP} covering a mobile user. The user has full \gls{6DoF} mobility, and receives video frames from the \gls{AP} at a fixed rate. The \gls{AP} uses a planar antenna array, capable of 3D beamforming. Through \gls{SLS} beamforming scheduled in either the \gls{A-BFT} or \gls{DTI}, the beam selection from the \gls{AP}'s codebook is kept up to date. For user-side beamforming, the \gls{HMD} is also equipped with a planar array, and CoVRage is integrated into it. As such, the \gls{HMD} does not require an explicit codebook for its antenna array, apart from one quasi-omnidirectional \gls{AWV}.
\subsection{Implementation}
In this subsection, we provide details on the different simulation components for the system evaluation, with an overview in Fig~\ref{fig:overview}.
\subsubsection{Network}
The \texttt{ns-3} discrete-event simulator handles network simulation. It simulates the wireless network from the physical up to the application layers, allowing for highly realistic simulation. Given \texttt{ns-3}'s modular nature, its functionality is easily extended. We use an existing module for IEEE 802.11ad and IEEE802.11ay support~\cite{ns3ad,ns3ay}. In addition, as part of this work, we developed a number of \texttt{ns-3} extensions, all made available to the community\footnote{https://github.com/orgs/XR-simulation/repositories}. Specifically, we provide the following:
\begin{itemize}
    \item \textbf{CoVRage}~\cite{covrage}: we integrate CoVRage into \texttt{ns-3}, maintaining compliance with the Wi-Fi specification. To this end, we equip the \gls{HMD} with a codebook containing a single directional sector. Whenever CoVRage is executed, its output \gls{AWV} overwrites the previous \gls{AWV} of the single sector, both for sending and receiving.
    \item \textbf{Retrain in A-BFT}: in the module, \gls{SLS} periods for beamforming with previously associated clients can only be scheduled in the \gls{DTI}. In our patch, such an \gls{SLS} can be scheduled during the \gls{A-BFT} as well. This is more efficient from a scheduling perspective, as this period is already reserved for beamforming. 
\end{itemize}
Furthermore, we integrate an additional traffic-generating application~\cite{burstGen} into this release of \texttt{ns-3}, as the simulator does not come with any traffic generators closely mimicking \gls{XR} traffic by default. This application generates an immediate data burst of a configurable size, at a fixed, configurable interval. This is the type of traffic expected with interactive \gls{XR}, with each burst corresponding to one video frame.

Furthermore, we use the module in conjunction with a quasi-deterministic channel model in Matlab, enabling highly realistic mmWave propagation based on ray tracing~\cite{qd}. We designed a $20 \times 10 \times$ \SI{10}{\meter} room in FreeCAD and input our own motion patterns (cf. Sec.~\ref{sec:user_motion}) into the channel model, but did otherwise not modify its code.
\subsubsection{Codebooks}
The \texttt{ns-3} module supports a highly realistic codebook format, in which each element's directivity, steering vectors for every (quantized) direction, and full \gls{AWV} of each sector can be provided through a config file. The module provides some example config files, but no generator for such files. To create our codebooks, we developed such a generator, based on the parser for these files. 

One challenging and underexplored aspect of codebook design is the design of the quasi-omni \gls{AWV}. For this task, we adapted an array synthesis script published by Mathworks~\cite{quasiOrig}. This script finds an \gls{AWV} using nonlinear optimization. Specifically, the \texttt{fmincon} function in Matlab is used to find an \gls{AWV} which minimizes the objective function $||\mathbf{b_a} - \mathbf{b_t}||_2$, where the vector $\mathbf{b_a}$ denotes the \gls{AWV}'s actual gains for a selection of predefined directions, and $\mathbf{b_t}$ denotes the target gains. The original work samples the gains on a per-degree basis in the azimuthal plane. For this work, we altered the script in several ways, available on this work's GitHub page. Firstly, the objective function was modified to $\max{(\mathbf{b_a})} - \min{(\mathbf{b_a})}$. By aiming to minimize the \textit{range} of gains, we can target a relatively constant gain in all directions without needing to explicitly define a target pattern $\mathbf{b_t}$. Next, we extended the script to support planar arrays, instead of only linear arrays. As this requires evaluation in both azimuth and elevation, rather than only azimuth, we uniformly sample \SI{1000}{} random azimuth-elevation combinations to determine $\mathbf{b_a}$. Furthermore, as fine-grained per-element amplitude control is not usually implemented in hardware, we restricted the \gls{AWV} to have fixed amplitudes, with only the phases being variable. With large arrays, the script's runtime was around one week.
\subsubsection{User motion}\label{sec:user_motion}
We subdivide the user's \gls{6DoF} motion into translational (i.e., walking) and rotational (i.e., turning one's body/head) motion respectively, and take the following approach to obtain these two datasets:
\begin{itemize}
    \item \textbf{Translational motion}: Through our previously published \texttt{pm4vr} mobility generator~\cite{pm4vr}, we generate synthetic user mobility paths. The simulator maps virtual paths, within a virtual experience, to physical paths actually walked by users. Through the use of Redirected Walking, users are subtly steered away from walls/obstacles in the physical space without noticeably influencing their intended virtual paths~\cite{rw}. The simulator generates a simple virtual path, randomly selecting one cardinal direction to move towards every step, then converts this to a physical path. The properties of the physical environment (size, shape, possible obstacles) are easily configured in the simulator, motivating its use.
    \item \textbf{Rotational motion}: As rotational input we simply use real traces of user rotation as measured on a modern \gls{HMD}. A test subject was asked to perform rotations of varying intensity while wearing a Meta Quest 2 \gls{HMD}. 
\end{itemize}

\section{Experiment Design}\label{sec:exp}
\begin{figure}[!t]
    \centering
    \includegraphics[width=\linewidth]{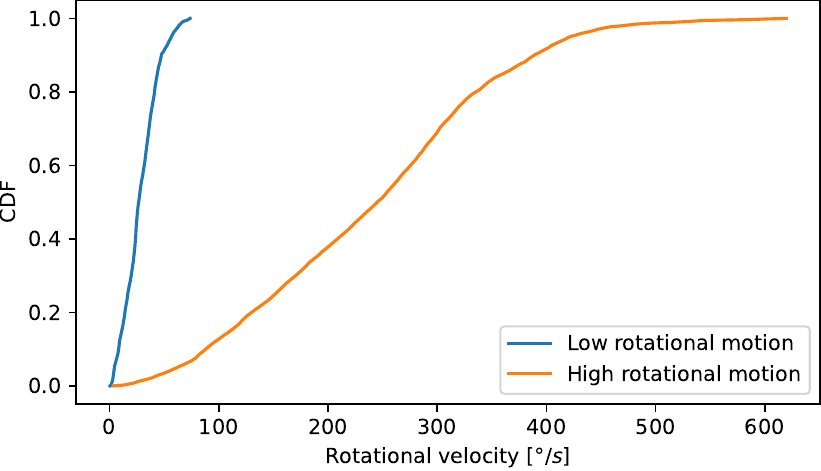}
    \caption{Rotational velocities, over a \SI{100}{\milli\second} window.}
    \label{fig:rotvel}
\end{figure}
The simulation experiments presented in this paper are focused on evaluating our end-to-end system, and its sensitivity to several network configuration parameters. 
\subsection{Parameters}
In each experiment, the user is placed at the center of the room, roaming around in it freely. For rotational motion, we use one of two collected trace files. In the \texttt{low} trace, the user was asked to perform slow, steady rotations, while in the \texttt{high} trace, the user performed frantic, rapid rotations as much as they were comfortable with while wearing an \gls{HMD}. Fig.~\ref{fig:rotvel} shows their rotational velocities. The application, generating \gls{XR}-like content, is set to a \SI{100}{\hertz} burst rate (i.e., one burst per \SI{10}{\milli\second}) and a total data rate of 2, 5, 7 or \SI{8}{Gbps} (with \SI{8}{Gbps} being the system's maximum PHY-level data rate). 
As a baseline for the CoVRage algorithm, the \gls{HMD} falls back to the regular sector-based system, or uses quasi-omni beams for all communications. When CoVRage is used, upcoming orientations are predicted through either simple constant-velocity extrapolation, or by using the Meta Quest 2's black-box predictor, which was constantly polled while collecting the rotational data traces. As a baseline, an oracle predictor, looking ahead in the trace, is used. The \gls{BI} is set to either \SI{102.4}{\milli\second} or \SI{1024}{\milli\second}. The former allows for more adaptive \gls{A-BFT} beamforming, while the latter reduces overhead. \gls{SLS} beamforming, always needed for \gls{AP} side transmit beamforming, may occur either during the \gls{A-BFT} or during the \gls{DTI}. In the first case, it is performed every \gls{A-BFT}, while in the latter it occurs at a regular interval of once every \SI{100}{\milli\second} or \SI{1000}{\milli\second}, being postponed until the start of the next \gls{DTI} if it is triggered during a \gls{BHI}. The \gls{MCS} is set to the maximum value of 21. We avoid employing an \gls{MCS} adaptation algorithm, as these are implementation-dependent, and underestimation of achievable \gls{MCS} may reduce performance, which would make results more difficult to interpret. Simulation time is limited to \SI{20}{\second} per simulation, keeping the runtime of the hundreds of configurations feasible.

The \gls{AP} is equipped with an $8 \times 8$ antenna array, providing a fair trade-off between signal strength, needed to achieve transmission at high \gls{MCS}, and beam width, ensuring the user does not leave the beam's range before the \gls{AP} can adapt. The array's sectors are aimed towards azimuths and elevations from the set $\{$\SI{-50}{\degree}, \SI{-30}{\degree}, \SI{-10}{\degree}, \SI{10}{\degree}, \SI{30}{\degree}, \SI{50}{\degree}$\}$ with 36 sectors covering every possible combination. A 37\textsuperscript{th} sector uses the quasi-omni \gls{AWV}. At the \gls{HMD}, we deploy a $64 \times 64$ array, enabling the CoVRage algorithm. For the sector-based baseline, we have to reduce this to $8 \times 8$, as very large arrays require many sectors, which is incompatible with the \gls{SLS} system, as its runtime scales linearly in the number of sectors. The quasi-omni baseline is run with each of the two arrays, enabling a fair comparison. These two baselines are the approaches currently used by commercial \gls{mmWave} solutions.  Table~\ref{tab:params} summarizes these parameters, with defaults in bold.

\begin{table}[!t]
    \caption{Simulation parameters (bold indicates default)\label{tab:params}}
    \centering
    \begin{tabular}{r|l}
    Parameter & Values\\
    \hline
    \hline
    Simulation time & \SI{20}{\second} \\
    \hline
    Room size & $10 \times $\SI{20}{\metre}\\
    \hline
    Rotational motion& Low, \textbf{High} \\
    \hline
    Data rate & 2,\textbf{5},7,\SI{8}{Gbps} \\
    Frame rate & \SI{100}{\hertz} \\
    \hline
    RX Beamforming & CoVRage \\
    RX Beamforming (baselines) & Quasi-omni only, sector-based\\
    Motion prediction & Extrapolation, \textbf{on-device} \\
    Motion prediction (baseline) & Oracle \\
    \hline
    \gls{BI} & \textbf{102.4}, \SI{1024}{\milli\second} \\
    Beamforming timing & in \gls{A-BFT}, in \textbf{\gls{DTI}} \\
    Beamforming frequency (\gls{A-BFT}) & once per \gls{BI} \\
    Beamforming frequency (\gls{DTI}) & \textbf{100}, \SI{1000}{\milli\second} \\
    \hline
    \gls{AP} array size & $8 \times 8$ \\
    \multirow{2}*{\gls{HMD} array size} & $8 \times 8$ (sectors, quasi-omni) \\
     & $64 \times 64$ (CoVRage, quasi-omni) \\
    \end{tabular}
    \end{table}

\subsection{Metrics}
For a high-\gls{QoE} \gls{XR} experience, full video frames need to consistently be delivered on-time. Even a small percentage of partially undelivered video frames can lead to a noticeable degradation in quality. In this work, we consider a video frame to be delivered successfully if and only if all packets comprising it arrive within a deadline of \SI{20}{\milli\second}, necessary to avoid motion sickness~\cite{mmwavevroverview}. For every successful delivery, we calculate its latency as the time between the video frame being offered to the transmitting application, and its final packet arriving at the receiving application layer. As a primary means of performance evaluation, we provide \glspl{CDF} of the video frame latency, from which one can read the percentage of successfully delivered video frames within any latency deadline. For each experiment with nonzero video frame loss, its reliability (i.e., the percentage of video frames fully delivered within \SI{20}{\milli\second}) is displayed on the plot.

\section{Results and Discussion}\label{sec:eval}
In this section, we present and discuss the results, and derive generic insights from these findings, provided in bold.

\begin{figure*}
    \centering
    \subfloat[\SI{5}{Gbps}]{\includegraphics[width=0.315\textwidth]{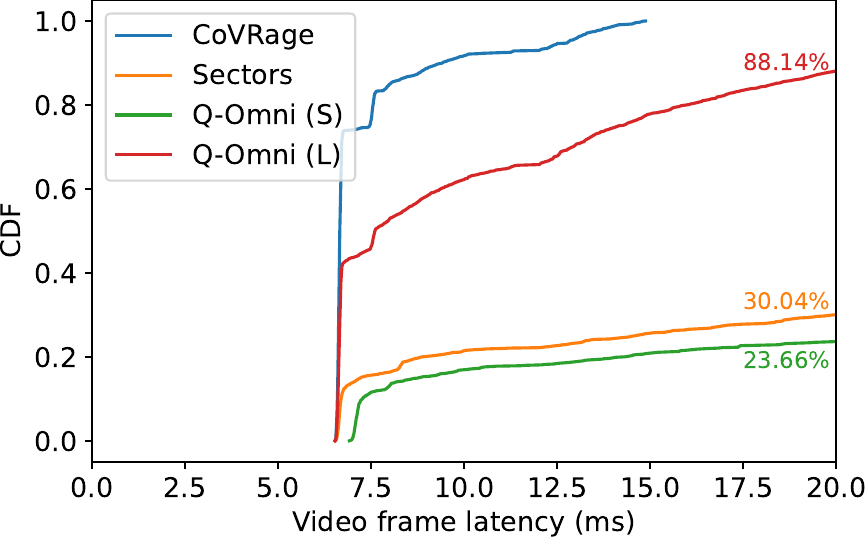}%
    \label{fig:bf_5000_High}}
    \hfill
    \subfloat[\SI{7}{Gbps}]{\includegraphics[width=0.315\textwidth]{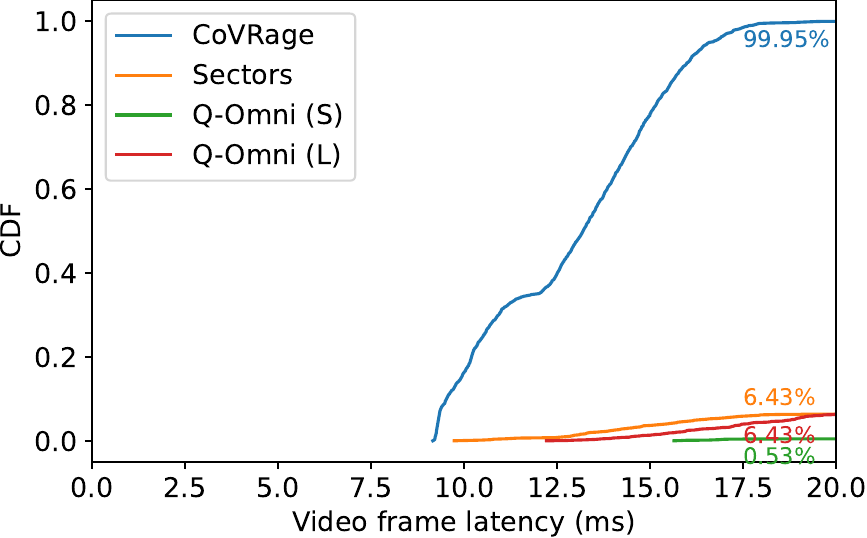}%
    \label{fig:bf_7000_High}}
    \hfill
    \subfloat[\SI{2}{Gbps}]{\includegraphics[width=0.315\textwidth]{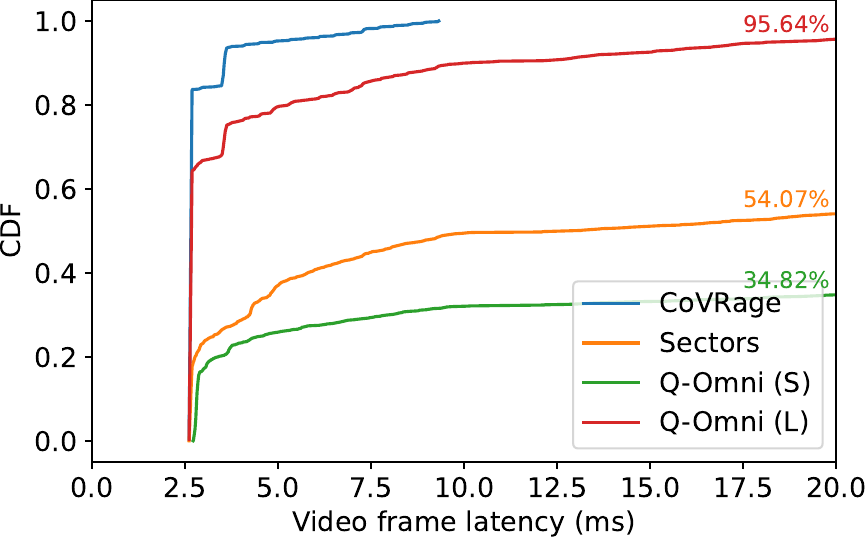}%
    \label{fig:bf_2000_High}}
\caption{Latency for CoVRage vs beamforming baselines at different data rates}
\end{figure*}

We begin by evaluating the performance of CoVRage compared to the sector and quasi-omni baselines. As Fig.~\ref{fig:bf_5000_High} shows, only CoVRage is able to maintain a stable connection, with all video frames arriving within \SI{15}{\milli\second}, well under the \SI{20}{\milli\second} deadline. With the best-performing baseline, \SI{11.9}{\percent} of video frames are lost, while the other two each lose over \SI{70.0}{\percent} of video frames. Minimum latency is \SI{6.5}{\milli\second}, indicating that, under perfect conditions, two-thirds of the \SI{10}{\milli\second} interval between two video frames is needed to transmit a single video frame~\cite{towards}. Intuitively, this indicates that the maximal throughput, under ideal conditions, is around \SI{7.5}{Gbps}. To evaluate this, we repeat the experiment with \SI{7}{Gbps} in Fig.~\ref{fig:bf_7000_High}. Here, CoVRage loses only \SI{0.05}{\percent}, with baselines losing at least \SI{93.5}{\percent}. At \SI{8}{Gbps}, loss is over \SI{99.8}{\percent} in all cases including CoVRage. This is expected, as maximal PHY-level speed is only \SI{8085}{Mbps}, which is unable to accommodate the \SI{8}{Gbps} stream plus overheads from four layers. If we instead reduce the data rate to \SI{2}{Gbps}, in Fig.~\ref{fig:bf_2000_High}, losses without CoVRage remain too high for a high-\gls{QoE} experience. This indicates that \textbf{beamforming approaches used on current-day general-purpose \gls{mmWave} devices only suffice for static \gls{XR} experiences, with minimal user rotation}.

\begin{figure*}
    \centering
    \subfloat[Different prediction approaches, \SI{7}{Gbps}]{\includegraphics[width=0.315\textwidth]{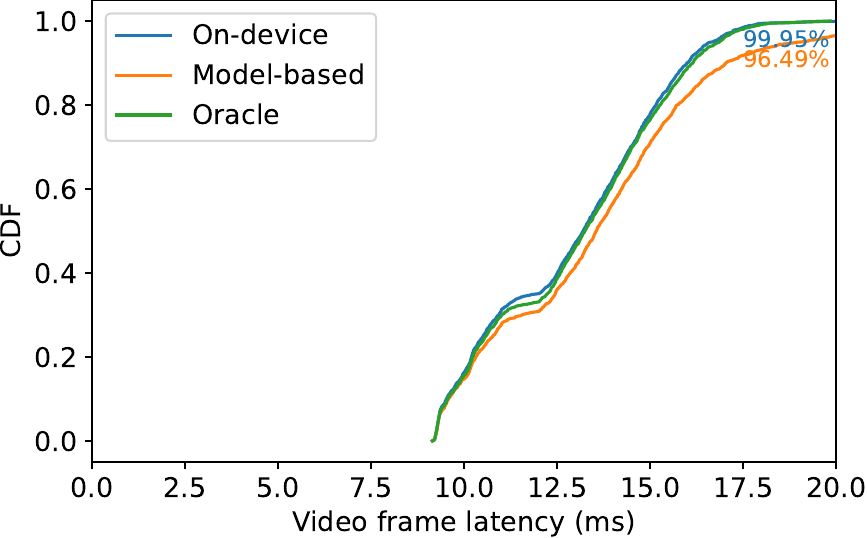}%
    \label{fig:pred_7000_High}}
    \hfill%
    \subfloat[\texttt{high} vs \texttt{low} motion, \SI{5}{Gbps}]{\includegraphics[width=0.315\textwidth]{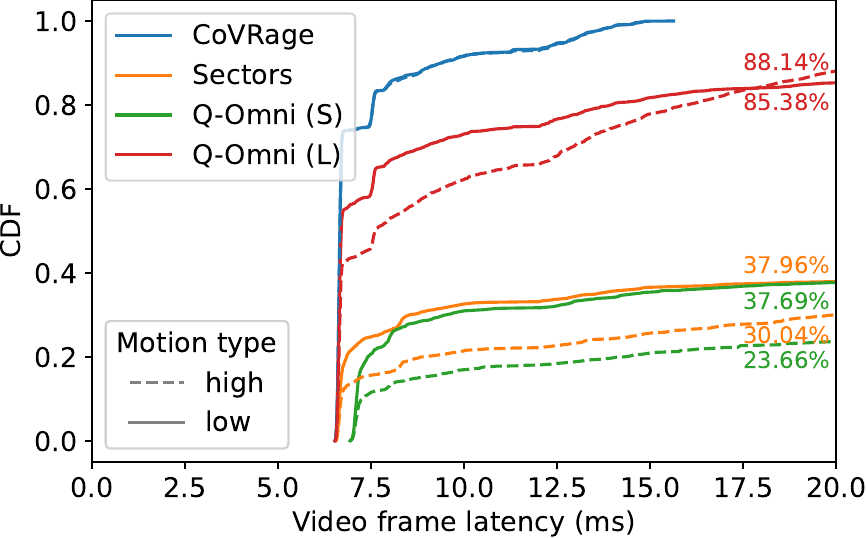}%
    \label{fig:motion_5000_both}}
    \hfill%
    \subfloat[Beamforming in \gls{DTI} or \gls{A-BFT}, different beamforming and beaconing intervals, \SI{5}{Gbps}]{\includegraphics[width=0.315\textwidth]{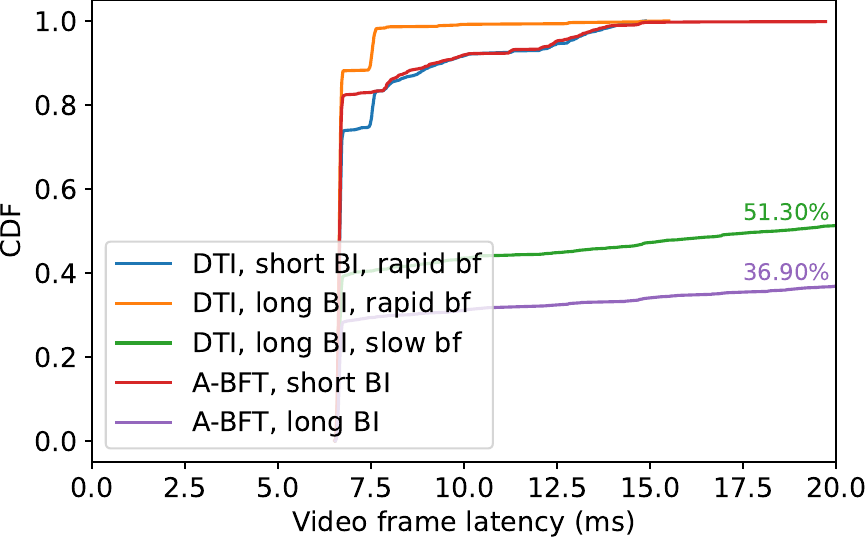}%
    \label{fig:interval_5000}}
    \caption{Further latency evaluations using CoVRage}
\end{figure*}

Next, we investigate the impact of the motion prediction approach on CoVRage, in Fig.~\ref{fig:pred_7000_High}, at \SI{7}{Gbps}. Performance is very similar in all three cases, with the on-device prediction nearly matching the oracle, and model-based being very slightly worse. On-device outperforming model-based is expected, as rotation prediction on even decade-old Oculus hardware was based on a constant-acceleration model, which they showed to outperform our simple constant-velocity model~\cite{RiftPrediction}. Furthermore, on-device prediction has access to more lower-level sensor readings. It is fair to assume that any modern \gls{HMD} provides such predictions, as they are needed for content generation. At lower data rates, the differences became nearly imperceptible.
For the next experiment, we repeat the first experiment, but with the \texttt{low} motion pattern, in Fig.~\ref{fig:motion_5000_both}. For the baselines, the intensity of the motion has a clear impact on performance, while for CoVRage the impact is negligible. This shows how CoVRage manages to synthesize effective beams during motion, regardless of motion intensity. \textbf{The CoVRage beamforming algorithm facilitates uninterrupted high-gain links to \gls{XR} devices, even during very rapid rotational motion}. To implement the algorithm, \textbf{on-device pose prediction is expected to be sufficiently accurate}.

Finally, we investigate the latency impact of reducing the duration of overhead periods, which do not allow for data transmission. By default, 10 \glspl{BI} occur each second, along with 10 \glspl{SLS} during the \gls{DTI}. We reduce overhead by reducing the \gls{BI} frequency by a factor 10, or by moving the \gls{SLS} to the \gls{A-BFT}. Note that, when applying both optimizations, the beamforming frequency is also reduced by a factor 10 as an effect. To more thoroughly evaluate the impact of these changes, we also consider the case where \gls{DTI} beamforming is triggered only once per second. Fig.~\ref{fig:interval_5000} shows latency for beamforming in \gls{DTI} and \gls{A-BFT}, with a \SI{102.4}{\milli\second} (``short'') or \SI{1024}{\milli\second} (``long'') \gls{BI}, and beamforming every \SI{100}{\milli\second} (``rapid'') or \SI{1000}{\milli\second} (``slow''). This shows how \textbf{rapid beamforming in the \gls{DTI} combined with a long \gls{BI} offers the best performance for interactive \gls{XR}}. This is the expected result, as increasing the \gls{BI} duration reduces the \gls{BHI}'s negative impact on video frame latency. We then compare this scenario to scenarios with a short \gls{BI} and rapid beamforming, either in \gls{DTI} or \gls{A-BFT}. In each case, the long tail of the \gls{CDF} starts at the \SI{7.6}{\milli\second} mark, but covers a significantly larger fraction of video frames with a short \gls{BI} than it does with a long \gls{BI}. This again makes sense, as 10 times as many \glspl{BHI} occur with a short \gls{BI}, making increased latency from waiting for a \gls{BHI} to finish significantly more likely in these cases. The \glspl{CDF} for \gls{DTI} beamforming exhibit a clear plateau between 6.75 and \SI{7.50}{\milli\second}, mostly absent in case of \gls{A-BFT} beamforming. This is caused by the \gls{SLS}, taking \SI{0.75}{\milli\second}, being scheduled arbitrarily inside the \gls{DTI}. Apart from this plateau, \gls{A-BFT} and \gls{DTI} beamforming perform almost identically at the same interval. Finally, increasing this beamforming interval has a drastic effect on performance, with around half to two thirds of video frames being lost. A user's orientation can change drastically within a second, and more frequent CoVRage beamforming is clearly needed to keep up with these changes. This shows how CoVRage achieves its intended result.  
   
\section{Conclusions and Future Work}\label{sec:conclusions}
In this work, we presented an end-to-end system approach for high-\gls{QoE} mobile wireless interactive \gls{XR} using \gls{mmWave}, evaluating its performance through simulations using realistic input data and channel models. Analyzing a wide array of scenarios revealed that, even at relatively modest data rates and motion patterns, regular beamforming approaches do not achieve acceptable \gls{QoE}. Through rapid head rotations, beams become misaligned before the system can adapt. With the \gls{HMD}-focused proactive beamforming algorithm CoVRage, the system was able to consistently deliver a data stream within \SI{20}{\milli\second} to a highly mobile user, even at \SI{7}{Gbps}, near the theoretical throughput limit. Based on these findings, we formulated a set of generic guidelines and insights regarding \gls{mmWave}-powered interactive \gls{XR}.

We also identify a number of potential extensions to this work. While the channel model and user mobility already provide a highly realistic single-user environment, real-world deployments are expected to often include multiple users. This would require the addition of more \glspl{AP} to satisfy the overall data rate need. The potential for spatial sharing is high with \gls{mmWave}, given its directional nature. However, \gls{AP} assignment and low-latency handovers in such a scenario are challenging, and is the focus of ongoing research. Furthermore, fully gauging the feasibility of this system will require an analysis of scalability, cost and energy consumption. We intend to extend our work with additional metrics in multi-user and multi-\gls{AP} scenarios, developing solutions to jointly optimize beamforming and \gls{AP} assignment, as to further assess the feasibility of future wireless \gls{XR} deployments. Furthermore, we intend to perform a small-scale version of these experiments with actual hardware, further validating our findings.
\section*{Acknowledgment}
This research was partially funded by the ICON project INTERACT and Research Foundation - Flanders (FWO) project WaveVR (Grant number G034322N). INTERACT was realized in collaboration with imec, with project support from VLAIO (Flanders Innovation and Entrepreneurship). Project partners are imec, Rhinox, Pharrowtech, Dekimo and TEO. The work of Filip Lemic was supported by the Spanish Ministry of Economic Affairs and Digital Transformation and the European Union – NextGeneration EU, in the framework of the Recovery Plan, Transformation and Resilience (Call UNICO I+D 5G 2021, ref. number TSI-063000-2021-6); the European Union's Horizon Europe's research and innovation programme under grant agreement nº 101139161 — INSTINCT project; and MCIN / AEI / 10.13039 / 501100011033 / FEDER / UE HoloMit 2.0 project (nr. PID2021-126551OB-C21). We thank Hany Assasa for support on the \texttt{ns-3} module and quasi-deterministic channel model. 

\bibliographystyle{IEEEtran}
\bibliography{IEEEabrv,bibliography}

\end{document}